\documentclass{article}
\usepackage{slashed,amsmath,amssymb,enumitem}
\usepackage[papersize={8.5in,11in}]{geometry}
\usepackage{graphicx}
\begin{document}
\title{Bimetric Relativity and the Opera Neutrino Experiment}
\author{J. W. Moffat\\~\\
Perimeter Institute for Theoretical Physics, Waterloo, Ontario N2L 2Y5, Canada\\
and\\
Department of Physics and Astronomy, University of Waterloo, Waterloo,\\
Ontario N2L 3G1, Canada}
\maketitle
\begin{abstract}
We investigate the possibility of explaining the propagation of neutrinos measured by the OPERA experiment with $\delta v_\nu=(v_\nu-c_0)/c_0$, where $c_0$ is the speed of light in vacuum, using a bimetric relativity model. The geometry of the bimetric model has two metrics in spacetime. One metric $g_{\mu\nu}$ possesses a null cone along which massless gravitons and photons travel with the constant speed $c_0$, while the other `matter' metric
${\hat g}_{\mu\nu}=g_{\mu\nu}+\beta\psi_\mu\psi_\nu$ has a null cone with a bigger speed $c > c_0$ with $0 < \delta v_\nu\ll c/c_0$. This second cone structure of spacetime prevents the neutrinos from being superluminal tachyons $v_\nu < c$. Such superluminal neutrinos would lead to severe depletion of neutrino energy, refuting the OPERA result. The charge-current source of the background gauge field $\psi_\mu$ is assumed to be baryon charge and the strength of the field $\psi_\mu$ is distance dependent, explaining the observation that for supernova SN1987a $v_\nu\simeq c_0$. Altering the path of the neutrinos through the earth or performing a space experiment can test the OPERA neutrino experimental result.
\end{abstract}

\section{Introduction}

The OPERA collaboration has reported that it has measured a superluminal speed of muon neutrinos~\cite{Opera}. The CNGS beam of muon neutrinos with mean energy 17.5 GeV with an energy spread about 50 GeV travels from CERN to the OPERA detector in the Gran Sasso Laboratory, a distance of 730 km. The OPERA group claims that their experiment gives a travel time for the ultrarelativistic neutrinos that is about 60 ns less than expected when compared to the speed of light. This suggests that neutrinos propagate with superluminal speed with $\delta v_\nu=(v_\nu-c_0)/c_0\sim 2.5\times 10^{-5}$, where $c_0$ is the speed of light in vacuum. This is a significant discrepancy of order 7.3 km/sec compared to the experimentally measured speed of light. The OPERA result can be expressed in terms of the parameter
$\eta=v_\nu^2/c_0^2-1\sim 2\delta v_\nu$, and for OPERA $\eta=5\times 10^{-5}$. The result is compatible with earlier experiments of high energy neutrinos e.g., MINOS~\cite{Minos} which gave $\eta=10.2\pm5.8\times 10^{-5}$. It is in serious discord with the observations of $\sim 10$ MeV neutrinos from supernova SN1987a, which yield the constraint $\delta < 4\times 10^{-9}$~\cite{SN}. Oscillating neutrino experiments provide significant constraints, which would require that if the OPERA result is correct, then the superluminal speed of neutrinos must apply to all three flavors of neutrinos~\cite{Coleman}.

We can consider superluminal tachyons as obeying standard Lorentz invariant transformation laws between inertial frames. If we write for $v_\nu > c_0$ that the relative speed between inertial frames is $v_\nu=iu_\nu$, then the Lorentz dilation factor $\gamma=1/(1-v^2/c_0^2)$ becomes $\gamma_{\rm tach}=1/(1+u^2/c_0^2)$ and the energy $E$ and rest energy $E_0=m_0c_0^2$ remain real quantities. However, now a particle is accelerated to ever increasing speeds as $E\rightarrow 0$ and this signals a severe instability of physics and the vacuum. As has been pointed out by Cohen and Glashow~\cite{Glashow}, this would lead to serious disagreement with particle physics phenomena. For example, the energy depletion due to tachyon Cherenkov-type bremsstrahlung radiation $\nu_\mu\rightarrow \nu_\mu+e^++e^-$ would predict that few, if any, neutrinos would reach the Gran Sasso detector with energies in excess of 12.5 GeV. This result cannot be reconciled with the OPERA claim for superluminal neutrino speed measurement. Further stringent constraints from the IceCube collaboration for shower energies 16 GeV~\cite{Icecube} yield $\eta < 4\times 10^{-10}$ and for energies of order 100 TeV they report $\eta < 2\times 10^{-11}$.

In view of the experimental evidence refuting tachyon neutrinos~\cite{Glashow,Ellis}, we may simply dismiss the OPERA claim of superluminal neutrinos. However, the OPERA result is so remarkable that we are tempted into studying whether there is a theoretical possibility to explain the OPERA result without violating the constraints of known particle physics and astrophysical experiments. In the following, we appeal to a model of spacetime for which local special relativity is violated~\cite{Moffat} by the introduction of two metrics, a metric $g_{\mu\nu}$ along the light cone of which gravitons and photons travel with the speed $c_0$, and a second ``matter'' metric ${\hat g}_{\mu\nu}$ in which massive particles can be influenced by a speed $c > c_0$. In this picture of spacetime, matter particles never exceed the speed $c$ and become tachyonic.

\section{Bimetric Geometry and Field Equations}

We shall construct a model in which spacetime has two metrics and two light cones. We introduce a vector field $\psi_\mu$ which relates the gravitational and electromagnetic metric $g_{\mu\nu}$ and the matter metric $\hat{g}_{\mu\nu}$ by~\cite{Moffat}
\begin{equation}
\label{eq:metrics}
{\hat g}_{\mu\nu}=g_{\mu\nu}+\beta\psi_\mu\psi_\nu,
\end{equation}
where $\beta>0$ is a dimensionless constant. The class of models
we consider is described by the action
\begin{equation}
\label{Totalaction}
S_{\rm TOT}=S_{\rm GR}[g]+S_\psi[\psi,g] +S_{\rm EM}[g,A]+S_{\rm
M}[\hat{g},{\rm matter\, fields}],
\end{equation}
where
\begin{equation}
\label{eq:GR action}
S_{\rm GR}[g]=-\frac{1}{\kappa}\int d^4x\sqrt{-g}(R[g]-2\Lambda),
\end{equation}
is the usual Einstein-Hilbert action, $\kappa=16\pi G/c_0^4$ and $\Lambda$ is the cosmological constant. The idea here is to present a model that embodies the physical content of a ``varying speed of light'' in a diffeomorphism invariant manner.

We assume a Proca action for the vector field $\psi_\mu$:
\begin{equation}
S_{\psi}[\psi,g]=\frac{1}{\kappa}\int
d^4x\sqrt{-g}\Biggl(-\frac{1}{4}g^{\mu\nu}g^{\alpha\beta}B_{\mu\alpha}B_{\nu\beta}+\frac{1}{2}m^2\psi^2\Biggr),
\end{equation}
where $B_{\mu\nu}=\partial_\mu\psi_\nu-\partial_\nu\psi_\mu$, and $\psi^2=g^{\mu\nu}\psi_\mu\psi_\nu$. Moreover, the electromagnetic field action is
\begin{equation}
S_{\rm EM}[g,A]=-\int d^4x\sqrt{-g}\biggl(\frac{1}{4}g^{\mu\nu}g^{\alpha\beta}F_{\mu\alpha}F_{\nu\beta}\biggr),
\end{equation}
where $F_{\mu\nu}=\partial_\mu A_\nu-\partial_\nu A_\mu$.

The fact that photons travel along the light cone $g_{\mu\nu}v^\mu v^\nu=0$ guarantees that Maxwell's equations retain their conformal invariance and that we always measure by electromagnetic experiments the standard speed of light $c_0$ in vacuum. We assume that {\it massless gravitons and photons} travel along the null cone determined by $g_{\mu\nu}v^\mu v^\nu=0$. Matter particles in the metric $g_{\mu\nu}$ obey the condition $v \leq c_0$.  When the gauge field $\psi_\mu$ becomes non-zero, then particles with non-vanishing mass can now have $c_0 < v < c$ with the speed $c$ determined by the null cone ${\hat g}_{\mu\nu}v^\mu v^\nu=0$. The gravitational metric field, the vector field $\psi_\mu$ propagate on the geometry described by $g_{\mu\nu}$, whereas all other matter fields will propagate on the geometry described by $\hat{g}_{\mu\nu}$. Thus if we consider the motion of a non-gravitational matter test particle, we assume that it is the geodesics of $\hat{g}_{\mu\nu}$ that are of physical interest.

We assume that the matter field action is one of the standard forms, but constructed out of
$\hat{g}_{\mu\nu}$, and therefore the field equations guarantee that the conservation laws
\begin{equation}
\hat{\nabla}_\nu
T^{\mu\nu}_{\rm M}[\hat{g}]=0,
\end{equation}
are satisfied where $\hat{\nabla}_\nu$ denotes the covariant derivative with respect to the $\hat{g}_{\mu\nu}$ metric connection, and
\begin{equation}
T^{\mu\nu}_{\rm M}[\hat{g}]=
\frac{2}{\sqrt{-\hat{g}}}\hat{g}^{\mu\alpha}\hat{g}^{\nu\beta}
\biggl(\frac{\delta S_{\rm M}[\hat{g}]}
{\delta\hat{g}^{\alpha\beta}}\biggr),
\end{equation}
are satisfied. \footnote{For a $U(1)$ gauge vector field $\psi_\mu$, we could invoke a Stuekelberg formalism that would make the action $S_{\psi}[\psi,g]$ gauge invariant and the quantized theory for the Proca action renormalizable, if we assume that the gravitational metric $g_{\mu\nu}=\eta_{\mu\nu}$, where $\eta_{\mu\nu}$ is the Minkowski metric of spacetime~\cite{Stuekelberg}.}

We could have considered a model in which the matter metric ${\hat g}_{\mu\nu}$ is given by~\cite{Moffat}:
\begin{equation}
{\hat g}_{\mu\nu}=g_{\mu\nu}+\beta\partial_\mu\phi\partial_\nu\phi,
\end{equation}
where $\phi$ is a scalar field. We shall consider for the present the bimetric model based on the gauge field $\psi_\mu$, for vector gauge fields exist in the standard model of particles.

Variation of~(\ref{Totalaction}) with respect
to $g_{\mu\nu}$ and $\psi_\mu$ leads to the field equations:
\begin{equation}
\label{eq:GR FEQ} \sqrt{-g}(G^{\mu\nu}[g]-\Lambda
g^{\mu\nu})=\frac{1}{2}\sqrt{-g}T_B^{\mu\nu}[g,\psi]+\frac{1}{2}\sqrt{-g}T_{\rm EM}^{\mu\nu}[g,A]
+\frac{\kappa}{2}\sqrt{-\hat{g}}T^{\mu\nu}_{\rm M}[\hat{g}],\\
\end{equation}
\begin{equation}
\label{VectorFEQ} \sqrt{-g}\Bigl(-\nabla_\nu
B^{\mu\nu}+m^2\psi^\mu \Bigr) =\beta\kappa
\sqrt{-\hat{g}}T^{\mu\nu}_{\rm M}[\hat{g}]\psi_\nu,
\end{equation}
where $\nabla_\nu$ denotes the covariant derivative formed from
the $g_{\mu\nu}$ metric connection. Moreover,
\begin{equation}
T_{B\mu\nu}=-B_{\mu\alpha}{B_\nu}^\alpha+\frac{1}{4}g_{\mu\nu}B^2
  +m^2\psi_\mu \psi_\nu -\frac{1}{2}g_{\mu\nu}
  m^2\psi^2,
\end{equation}
and
\begin{equation}
T_{\rm EM}^{\alpha\beta}=F^{\alpha\mu}{F^\beta}_\mu-\frac{1}{4}g^{\alpha\beta}F^2.
\end{equation}

It can be shown that the field equations and matter conservation laws are consistent with the Bianchi
identities.

The field equations~(\ref{VectorFEQ}) have the important property that $\psi_\mu=0$ is always a solution regardless of the
matter content of spacetime, in which case the conventional general relativity coupled to matter models are realized and there
is no conflict with experiment. In regions of spacetime where $\psi_\mu$ is nonvanishing and $g_{\mu\nu}v^\mu v^\nu > 0$, we can restrict
ourselves to frames that are aligned with the vector field: $\psi_\mu\rightarrow(1,0,0,0)$, and we have reduced the gauge
group of the orthonormal frames to $\mathrm{O}(3)$. Thus we see that $\psi_\mu\neq 0$ can play the role of a {\it vacuum condensate}
$\langle\psi_\mu\rangle_0$ that can be said to spontaneously `break' local Lorentz invariance~\cite{Moffat2}.

The model that we have introduced here is a ``metric theory of gravity'' in the sense that all fields with massive particles respond to the metric $\hat{g}_{\mu\nu}$. The dynamics that determine $\hat{g}_{\mu\nu}$ involve the tensor $g_{\mu\nu}$ as well as the vector $\psi_\mu$ and therefore preferred frame effects are possible. We expect that the vector field will essentially lead to repulsive effects (the presence of $\psi_\mu$ locally increases the speed of matter propagation, thereby effectively decreasing the gravitational coupling to the matter). The magnitude of the vector field is dependent upon the local matter energy density.

\section{Background Gauge Field and the OPERA Neutrino Speeds}
\label{sect:Fields}

The local special relativity metric is given by $g_{\mu\nu}=\eta_{\mu\nu}$ with
\begin{equation}
\label{SRM}
ds^2\equiv\eta_{\mu\nu}dx^\mu dx^\nu=c_0^2dt^2-(dx^i)^2,
\end{equation}
where ($i,j=1,2,3$) and we have used the metric signature ${\rm diag}(\eta_{\mu\nu})=(+1,-1,-1,-1)$. The matter metric is
\begin{equation}
\label{Mattermetric}
d{\hat s}^2\equiv {\hat g}_{\mu\nu}dx^\mu dx^\nu=(\eta_{\mu\nu}+\beta\psi_\mu\psi_\nu)dx^\mu dx^\nu.
\end{equation}
The latter can be written as
\begin{equation}
d{\hat s}^2=c_0^2dt^2\biggl(1+\beta\psi_0^2\biggr)-(\delta_{ij}-\beta\psi_i\psi_j)dx^idx^j.
\end{equation}
We have\footnote{In GR, with one local lightcone, we can always perform a diffeomorphism transformation to remove the time dependence of $c$. This corresponds to being able to choose units in which rigid ruler and clock measurements yield $\Delta c=0$.  In the bimetric relativity theory, we cannot simultaneously remove the time dependence of $c$ and $c_0$ by performing a diffeomorphism transformation. If we choose $c_0$ to be constant, then the time dependence of $c$ is non-trivially realized.}
\begin{equation}
c=c_0\biggl(1+\beta\psi_0^2\biggr)^{1/2}.
\end{equation}
We see that the speed $c$ in the matter metric is space and time dependent.

The null cone equation $ds^2=0$ describes gravitational and electromagnetic wave signals moving with the constant measured speed $c_0$, whereas $d{\hat s}^2=0$ cannot be satisfied along the same null cone lines, but determines an expanded null cone with the speed $c > c_0$.  The bimetric
null cone structure is described in Fig.1.
\vskip 0.2 in
\begin{center}
\includegraphics[width=3.0in,height=3.0in]{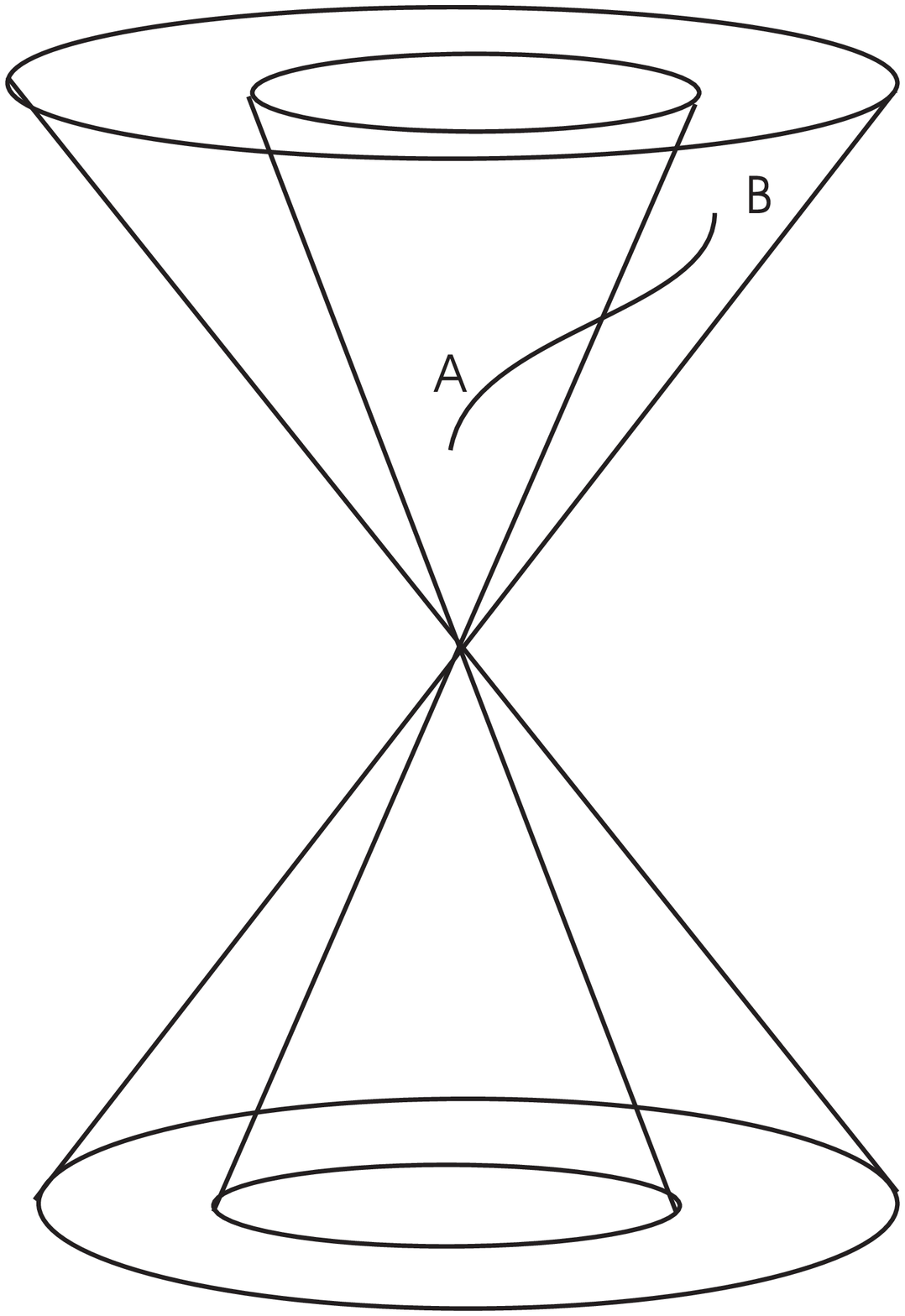}
\end{center}
\vskip
0.1 in
\begin{center}
Fig. 1. Bimetric light cones showing the timelike travel of neutrinos from A to B between the graviton-photon and matter metric light cones.
\end{center}
\vskip 0.1 true in

We postulate that the metric (\ref{SRM}) is invariant under {\it local} \footnote{We consider a small patch of spacetime in which $g_{\mu\nu}\approx \eta_{\mu\nu}$, so that we can restrict our attention to local Lorentz transformations.} Lorentz transformations
\begin{equation} x^{'\mu}={\Lambda^\mu}_\nu x^\nu,
\end{equation}
where ${\Lambda^\mu}_\nu$ are constant matrix coefficients which satisfy
the orthogonality condition
\begin{equation}
{\Lambda^\mu}_\nu{\Lambda_\mu}^\sigma={\delta_\nu}^\sigma.
\end{equation}
The matter metric ${\hat g}_{\mu\nu}$ is not invariant under local Lorentz transformations.

The `physical' matter metric ${\hat g}_{\mu\nu}$ describes the geometry in which standard model massive particles propagate and interact with a value of the speed $ c > c_0$. Because all particle matter fields are coupled to the same metric ${\hat g}_{\mu\nu}$ in the same way, the weak equivalence principle is satisfied in this metric frame. However, because ${\hat g}_{\mu\nu}$ does not couple to matter in the same way as in GR unless $\psi_\mu=0$, the strong equivalence principle is not satisfied.

We now consider the propagation of fermions in the matter metric spacetime (\ref{Mattermetric}) in the presence of the background gauge field $\psi_\mu$. Such models have been considered in the context of Lorentz violating models~\cite{Kostelecky}. The Lagrangian is given by
\begin{equation}
{\cal L}_f=\frac{1}{4}B^{\mu\nu}\biggl(1-\frac{\Delta}{M^2}\biggr)B_{\mu\nu}+\frac{1}{2}m^2\psi^2
+{\bar\psi}_f(i\gamma^\mu\partial_\mu-g_i\gamma^\mu\psi_\mu)-m_f{\bar\psi}_f\psi_f,
\end{equation}
where $\Delta=-\partial^i\partial_i$ and $M$ is a Lorentz violating mass scale. One may now have fermions with velocities $c_0 < v_f < c$ depending on the strength of the coupling constants $g_i$. We know that for the electron, $v_e/c_0\leq 1$ to a precision of at least $10^{-15}$~\cite{Coleman}. We assume that only the (vector-axialvector) coupling constant for neutrinos $g_\nu$ is sufficiently big to couple to the background field $\psi_\mu$. 

The equations of motion for $\psi_\mu$ are given by
\begin{equation}
\partial_\nu B^{\mu\nu}+m^2\psi^\mu=J_B^\mu,
\end{equation}
where
\begin{equation}
J^\mu_B=T^{\mu\nu}_{\rm Baryon}\psi_\nu.
\end{equation}
We have assumed that the source current is the baryon charge current $J_B^\mu$ and that the baryon charge $Q_B$ is
\begin{equation}
Q_B=\int d^3xJ_B^0.
\end{equation}
Here, $J_B^0$ is determined by the earth's baryon charge, $Q_B\sim q_BM_\oplus/m_p$, where $m_p$ is the mass of the proton.

By adopting the condition
\begin{equation}
\partial^\nu\psi_\nu=0
\end{equation}
the equations of motion become
\begin{equation}
{\vec{\nabla}}^2\psi_\mu -\frac{\partial^2\psi_\mu}{\partial t^2}+m^2\psi_\mu=J_{B\mu}.
\end{equation}
The static spherically symmetric point particle solution for $\psi_0$ is
\begin{equation}
\psi_0(r)=-q_B\frac{\exp(-mr)}{r},
\end{equation}
where $\ell=1/m$ is the distance scale that determines the magnitude of $\psi_0(r)$ as a function of radius $r$.

We can now reconcile the OPERA neutrino results with the SN1987a data constraints by requiring that $r_{\oplus}/\ell_{\rm Opera}\sim 1$, while for SN1987a, $r_{\rm SN}/\ell_{\rm Opera}\gg 1$, so that $\psi_0\approx 0$ and for neutrinos emitted from SN1987a, $v_\nu\sim c_0$.
An experimental test that can check the TOF of the neutrinos can now be proposed. Since the difference $\delta c=(c-c_0)/c_0$ depends on the magnitude of the background field $\psi_\mu$, and this magnitude is dependent on $r$, then directing the neutrino pulses through the earth instead across its surface would decrease the altitude of the neutrino pulses and increase the size of $\delta c$ and increase the muon neutrino velocity $v_\nu$.

\section{Concluding Remarks}

A bimetric theory is proposed in which two metrics are associated with spacetime.  Gravitons and photons travel along the the null cone of the metric $g_{\mu\nu}$ and for $\psi_\mu=0$ matter particles obey $v < c_0$. For a non-zero background gauge field $\psi_\mu$, the speed $c$ can exceed $c_0$ and matter particles traveling in the metric ${\hat g}_{\mu\nu}$ spacetime can now have $c_0 < v < c$.  The magnitude of $\psi_0$ determines the size $\delta c$ of the increase in the speed $c_0$. Superluminal tachyon neutrinos are prevented from existing, for the neutrino velocity $v_\nu < c$ and the Lorentz dilation factor becomes $\gamma=1/(1-v_\nu^2/c^2)$ in the metric ${\hat g}_{\mu\nu}$. The reason that the SN1987a data with its lower energy $\sim 10$ MeV is consistent with $v_\nu\sim c_0$ is due to the distance dependence of the gauge field strength $B_{\mu\nu}$. For the supernova SN1987a $\psi_0\sim 0$ and $\delta c\sim 0$. This suggests that by altering the altitude of the OPERA neutrino pulses through the earth a detectable distance effect could be measured for $\delta v_\nu$. A neutrino detector on a spacecraft or on the moon could also be utilized to test the distance dependence of $v_\nu$.

The striking experimental result reported by the OPERA experiment requires a speculative theoretical explanation requiring a violation of Lorentz invariance and possibly a novel way of describing the geometrical structure of spacetime. Experiments are obviously required to be performed to confirm the OPERA result and hopefully they can be carried out with sufficient precision in the not too distant future.

\section*{Acknowledgments}

I thank Viktor Toth and Martin Green for helpful discussions. This work was supported by the Natural Sciences and Engineering Research Council of Canada. Research at the Perimeter Institute for Theoretical Physics is supported by the Government of Canada through NSERC and by the Province of Ontario through the Ministry of Research and Innovation (MRI).


\end{document}